# Adsorption of SARS CoV-2 Spike Proteins on Various Functionalized Surfaces Correlates with the Strong Infectivity of Delta and Omicron Variants

*Daniela Dobrynin[1], Iryna Polishchuk[1], Lotan Portal[1], Ivan Zlotver[1], Alejandro Sosnik[1]\*, Boaz Pokroy[1] \**

[1] Department of Materials Science and Engineering, Technion – Israel Institute of Technology, 32000 Haifa, Israel.

## Abstract

The SARS CoV-2 virus emerged at the end of 2019 and rapidly developed several mutated variants, specifically the Delta and Omicron, which demonstrate higher infectivity and escalating infection cases worldwide. The dominant transmission pathway of this virus is via human-to-human contact and aerosols, but another possible route is through contact with surfaces contaminated with SARS-CoV-2, often exhibiting long-period survival. Here we compare the adsorption capacities of the S1 and S2 subunits of the spike (S) protein from the original variant to that of the S1 subunit from the Delta and Omicron variants. The results clearly show a significant difference in adsorption capacity between the different variants, as well as between the S1 and S2 subunits. Overall, our study demonstrates that while the Omicron variant is able to adsorb much more successfully than the Delta, both variants show enhanced adsorption capacity than the original strain. We also examined the influence of pH conditions on the adsorption ability of the S1 subunit and found that adsorption was strongest at pH 7.4, which is the physiological pH. The main conclusion of this study is that there is a strong correlation between the adsorption capacity and the infectivity of the various SARS CoV-2 variants.

## 1. Introduction

Coronaviruses are a highly diverse family of positive single-stranded RNA viruses, whose genome is packed within a capsid formed by the nucleocapsid (N) and surrounded by an envelope formed by the envelope protein (E).[1,2] Three structural proteins are associated with the envelope, namely the membrane protein (M) and E that assemble the virus and the spike protein (S) that mediates virus entry into a host cell.

In December 2019, a new human coronavirus, Severe Acute Respiratory Syndrome coronavirus 2 (SARS-CoV-2, COVID-19), was identified and declared a global pandemic by the World Health Organization (WHO).[3,4] Since then, the virus has continued to mutate and spread across the world, with almost 380 million cases and more than 5.6 million deaths in the last 2 years.

The spike of SARS-CoV-2 is a homotrimer glycoprotein named spike (S) protein composed of two subunits S1 and S2 that is responsible for the attachment and entry of the virus into the host cell via the host membrane receptor, angiotensin-converting enzyme 2 (ACE2).[5,6] The ACE2 receptor is ubiquitous and is distributed mainly in the lungs, heart, intestine, kidney and eyes.[7] A scheme of the general structure of SARS CoV-2 and its attachment and entry into the host cell are presented in **Figure 1A**. The S1 subunit consists of two domains, N- and C-terminal, and it binds the ACE2 receptor through the receptor-binding domain (RBD) located on the N-terminal domain. The S2 subunit, composed of five domains enclosing mainly hydrophobic amino acids, plays a key role in fusion of the virus with the host cell membrane.[8],[9] Such fusion and cell entry are both activated by transmembrane protease serine 2 (TMPRSS2), which cleaves the S protein in S1/S2 and S2 sites.[6,10,11] The transmissibility of respiratory viruses is governed by their infectivity, the contagiousness of the infected individual, the susceptibility of the exposed individual, the contact patterns between them, and environmental factors.[12] Therefore, any mutation in the S protein, and especially in its RBD, can affect the virus's infectivity and consequently its transmissibility. Recently yet another variant has been identified, namely the B.1.1.529 or Omicron variant, which exhibits 39 mutations in its S protein, including 15 in the RBD.[13,14] As a consequence, this new variant has even higher transmissibility characteristics than the Alpha, Beta, Gamma and the especially aggressive Delta, and has therefore become the dominant variant globally.[15,16]

One of the possible pathways by which SARS-CoV-2 spreads and transmits to humans is through contact with solid surfaces contaminated with the virus.[17–23] For instance, the virus has been shown to survive on surfaces such as plastics, fabrics, metals and glass, from minutes up to days.[19,24] Survival of the virus depends on environmental conditions such as temperature,[26, 27] humidity,[27] and exposure to light.[29,30] Furthermore, it was shown that while SARS-CoV-2 is viable for longer periods of time on surfaces that are surrounded by aqueous fluids, it also remains infectious in its dry form.[31,32] **Figure 1B** shows a 3D representation of hydrophilic and hydrophobic amino acids on the surface of the S protein of SARS-CoV-2, a distribution that most probably contributes to the formation of nonspecific interactions with the environment.

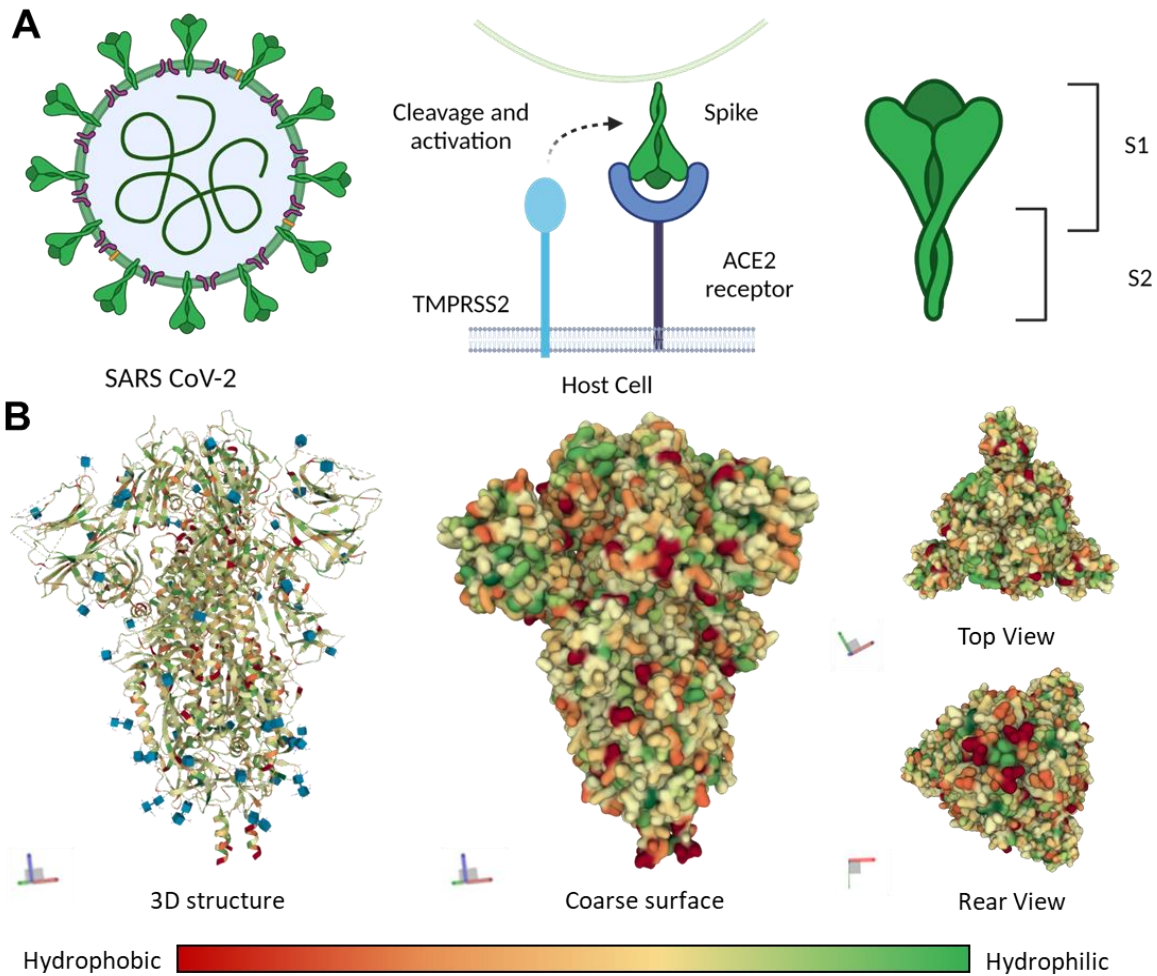

**Figure 1.** (A) Schematic presentation of SARS-CoV-2 and Spike protein attachment to the ACE2 receptor. (B) Spike 3D structure and presentation of the hydrophilicity-hydrophobicity of the surface: side, top and rear view of the original strain. (RCSB PDB: 6ZWV) https://www.rcsb.org/3d-view/6ZWV/1, created with BioRender.

Lately some doubts have been expressed about the ability of SARS-CoV-2 to infect humans via surfaces, but the issue is still under scientific debate.[32] Protein adsorption onto solid surfaces has been a hot topic in the biomedical field for decades. In particular, protein adsorption is a key step in the foreign-body response to implants in the extra-cardiovascular and cardiovascular systems[33] and various methods have been used to elucidate the mechanisms of attachment, the protein orientation on the surface, the conformational changes that take place at the surface, and the adsorbed amount.[34–36]

Characterization of the interaction of SARS-CoV-2 S1 and S2 subunits with different chemically modified surfaces is an essential stage in preventing further fomite transmission, and thus calls for the use of facile, easily accessible, and reproducible techniques. The Quartz Crystal microbalance (QCM) system comprises a disc-shaped quartz crystal that measures the deposited mass on the surface via the difference in resonant frequency of the quartz during the experiment.[38-41] It is a recognized research tool in surface engineering,[41,42] biophysics,[42] biomaterials science,[44,45] and electrochemistry.[45,46] The advantage of this method is that the surface of the crystal can be coated with materials displaying a broad spectrum of properties, and their adsorption on these surfaces can then be studied. The change in frequency correlates to the adsorbed mass density ($\Delta m$) on the active electrode area. For rigid and thin films, $\Delta m$ can be calculated from the Sauerbrey equation (1):[47]

$$\Delta m_{Sauerbrey} = -\frac{C}{n} * \Delta f \quad (1)$$

where $C$ is a constant (17.7 ng cm$^{-2}$ Hz$^{-1}$ for 5 MHz), $n$ is the overtone number, and $f$ is the frequency.

Rigid film refers to a film in which an acoustic wave propagates without any energy loss (elastically). Conversely, for viscoelastic and soft films it is possible to measure a dissipation shift (energy loss), which aids in understanding the viscoelastic properties of the adsorbed layer and its coupling to the crystal surface. The adhering layer can be characterized in terms of values of the thickness and the mass extracted at multiple frequencies, followed by application of a viscoelastic model (such as the Voigt model).[47,48] A striking advantage of this analytical tool is that it allows noninvasive, extremely sensitive measurements (changes in mass on the nanogram level), and real-time measurements.

In this work, we utilized dissipation monitoring QCM (QCM-D) to comparatively characterize the adsorption of the SARS CoV-2 S protein subunits S1 and S2 of the original strain and the S1 subunit of two more transmissible variants, namely Delta (S1-δ) and Omicron (S1-o), to gold-coated QCM-D sensor surfaces modified with various functionalized self-assembled monolayers (SAMs). This comparison allowed us to study the preferable adsorption of the S proteins, under different pH conditions, to various surface chemistries.

The development of SAMs on surfaces has proved enormously useful in many areas of science and technology.[49,50] Among the most extensively studied and utilized SAMs are alkanethiols, and to some extent also disulfides.[51] These compounds bind strongly to coinage and noble metals such as gold,[51–53] silver,[54–56] copper,[57,58] platinum,[59,60] palladium,[61,62] and mercury.[63,64] The binding energy of thiols to gold is ~40 Kcal/mol.[65] Other advantages of applying alkanethiols in our study are the ease of use and the wide range of possible chemical functionalities.

Overall, the results of this research provide important insight into the mechanism of binding of SARS CoV-2 to various surfaces.

## 2. Results

To investigate the affinities of SARS CoV-2 S-proteins to various surface functionalities, we chose to modify the gold-coated quartz crystals employed in the QCM-D with alkanethiols bearing different functional end groups. Following overnight immersion of the gold-coated sensors in a dilute solution of alkanethiol in ethanol we qualitatively studied, via water contact angle measurements, the formation of the various SAMs on the surface compared to the bare gold surface as a control. Each differently functionalized SAM produced a surface with a specific chemistry, as observed by the degree of hydrophilicity/hydrophobicity obtained at the surface. **Table 1** summarizes the water contact angle values of the variously functionalized surfaces.

**Table 1.** Structure of the different self-assembled monolayers used to modify the QCM gold sensor and the water contact angle after modification

| SAM | 11M1U | 11A1U | 11MUN | No SAM (bare gold surface) | TDT | FOT |
|---|---|---|---|---|---|---|
| End group | -OH | -NH$_2$ | -COOH | - | -CH$_3$ | -CF$_3$ |
| Water contact angle (°) | 48.0 ± 1.3 | 55.9 ± 0.5 | 64.7 ± 1.4 | 91.6 ± 3.9 | 110.1 ± 2.1 | 114.9 ± 1.1 |
| Water drop image | 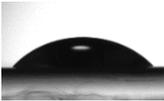 | 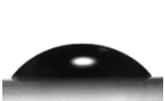 | 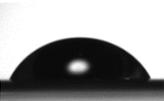 | 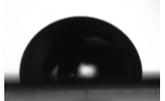 | 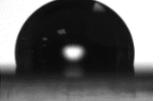 | 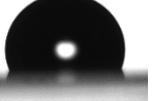 |

Next, we integrated the chemically modified sensors to study the adsorption of the S1 and S2 proteins. The experiments were performed utilizing the QCM-D tool while keeping a constant protein concentration of 0.05 mg mL$^{-1}$. **Figure 2** presents data on the frequency and dissipation shifts collected during the adsorption process. Upon addition of the protein solution we observed a sudden drop in the frequency values, accompanied by an increase in dissipation. The greater the amount of protein adsorbed, the more pronounced was the decrease in frequency. As a final step, we rinsed the sensor surface for 30 min with protein-free PBS buffer. This rinsing caused a small increase in the frequency and a small decrease in dissipation owing to detachment of unbound proteins. As can be observed from the table, the frequency and dissipation values did not revert to those observed prior to protein addition. This served as proof that most of the adsorbed proteins had remained tightly bound by Au-S bonding to the sensor surfaces.

**Figure 2** shows that each protein − SAM combination resulted in a different curve behavior indicating distinct protein affinity. In the case of the S1 protein, all surface modifications demonstrated a drop in frequency greater than that of the SAM-free surface. In particular, the strongest decrease in frequency was observed for the TDT, FOT, 11M1U and 11MUN surface modifications, while the frequency drop was intermediate for 11A1U and minimal for bare gold. In all of these cases, except for 11MUN and 11A1U SAMs the dissipation was rather high (**Figure 2B**).

Interestingly, comparison of the S1 subunit to that of the δ- and o-variants revealed a distinct difference, namely a major decrease in frequency and increase in dissipation curve, in the case of 11MUN SAM compared to other modifications (**Figure 2C,E**). Another interesting finding concerning the S1-δ subunit was that the decrease in its frequency was lowest on the surface functionalized with the hydrophilic 11M1U SAM. In addition, in the case of the S1-o subunit, during the last 30 min of the experiment (which included PBS-rinsing) there was no increase in the frequency and no decrease in the dissipation, as were observed for both the S1 and the S1-δ subunits (**Figure 3F**). This last finding might suggest that binding of the S1-o subunit on the various surfaces was stronger than that of the previous variants, which had exhibited some detachment during the rinsing cycle.

In the case of the S2 subunit, the most significant change in frequency indicated more adsorption to the hydrophobic TDT, while 11A1U and FOT showed the lowest adsorption, as expressed by the lowest frequency changes (**Figure 2G**).

In order to calculate the adsorbed protein mass, we needed to choose the most suitable model. To this end we ascertained whether the attached proteinous surface is thin or thick, rigid or soft. The criterion for thin film is that the $\Delta f$ (change in frequency during the adsorption process) is less than 2% of the initial resonance frequency.[47,48] In all our experiments this value was less than 0.1%, meaning that we could indeed apply the thin-film model to all of them. In cases where, in addition, the $\Delta D$ (dissipation change) was less than $1.5 \times 10^{-6}$ when the 3rd overtone was used (n = 3), we utilized the Sauerbrey Equation (1) to calculate the mass density of the protein layer adsorbed on the surface. When the $\Delta D$ was greater than $1.5 \times 10^{-6}$, we calculated the parameters of soft films more accurately by applying the Voigt model.

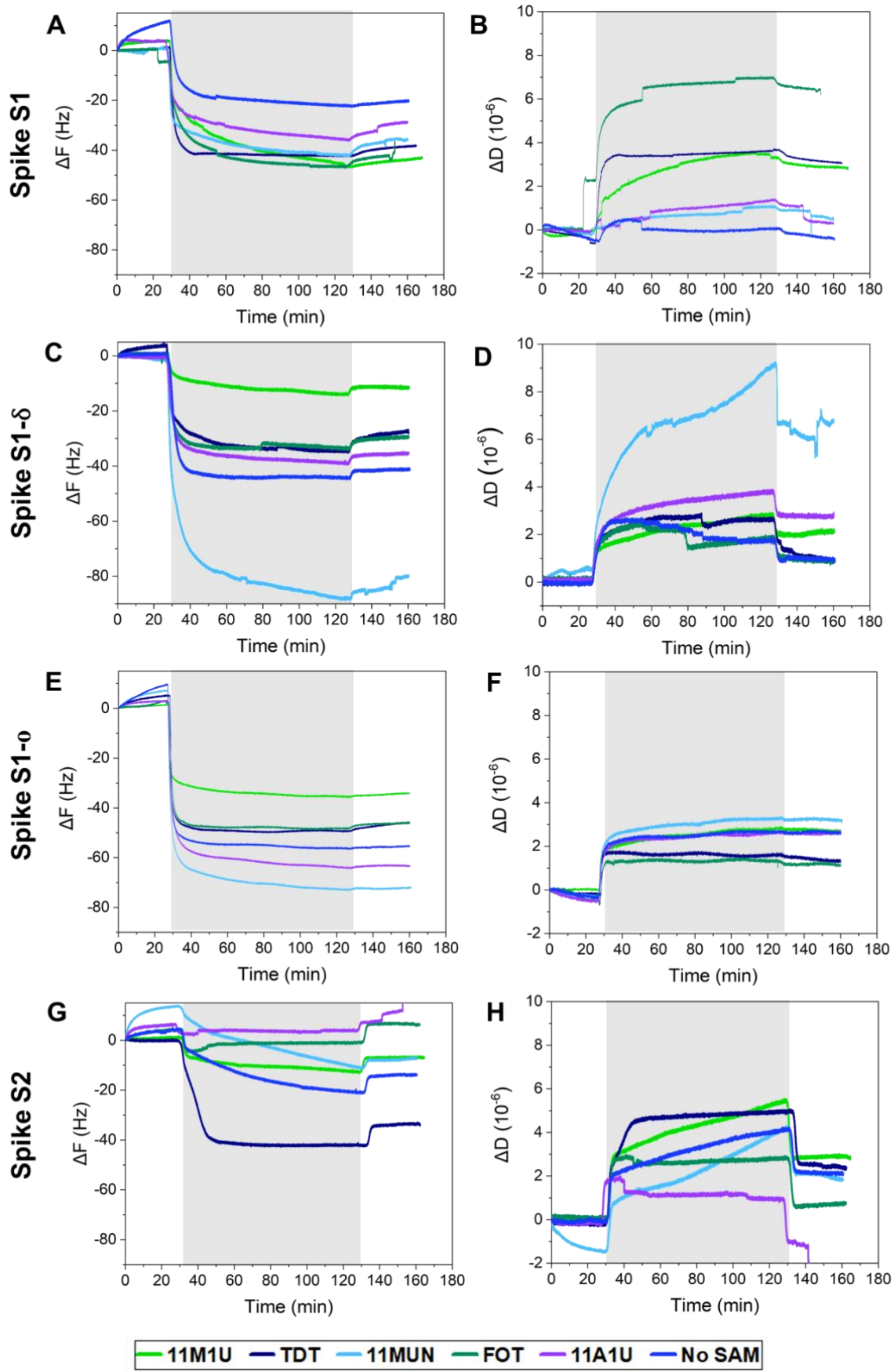

**Figure 2**. Real time adsorption curves monitored using QCM-D. The frequency and dissipation shifts over time of the Spike S1 (A,B), Spike S1-δ (C,D), Spike S1-o (E,F) and Spike S2 (G,H) proteins on various SAM-functionalized surfaces. The grey background represents the adsorption process of the added proteins (0.05 mg/mL) and the white background **represents protein-free PBS.** The 3$^{rd}$ overtone is presented (n = 3).

**Figure 3** presents the calculated mass density of the S1, S1-δ, S1-o and S2 subunits adsorbed onto the various chemically modified sensors.

By comparing the adsorption mass densities of S1 and S1-δ proteins (**Figure 3A**) we can establish different affinity trends. In the case of the S1 subunit, adsorption on the hydrophilic surfaces (11M1U, 11A1U and 11MUN) was favorable compared to that on the hydrophobic ones (TDT and no-SAM). The highest amount was adsorbed onto the surface functionalized with 11MUN (760 ng cm$^{-2}$) (**Figure 3A**). In contrast, the S1-δ subunit demonstrated an increase in adsorption to the hydrophobic surfaces (TDT and no-SAM) and a decrease to the hydrophilic ones (11A1U and 11M1U), except for 11MUN, which showed a ~2-fold increase in mass density compared to S1. The latter may also indicate that the adsorbed protein layer was a multilayer.[67]

The S1-o subunit exhibited the highest adsorption to hydrophobic surfaces (FOT, TDT and bare gold) relatively to all other S1 subunits. Additionally, it was found to adsorb better than the S1-δ subunit onto hydrophilic surfaces (11A1U and 11M1U), but slightly less well than the S1-δ subunit onto the negatively charged 11MUN.

We further calculated the average adsorbed mass density of the proteinous layers formed on all the studied functionalized surfaces in the case of each S1 subunit. From this simple calculation we found that the average mass density of the S1, S1-δ and S1-o subunits was 602, 660, and 891 ng/cm$^2$, respectively. This result implies that, on average, adsorption of the S1-o subunit on various surfaces was the highest.

In comparing the kinetics of the process of adsorption of each of the S1 subunits onto the 11MUN-functionalized surface during the first 5 min of the adsorption, we calculated that the changes in the frequency (in Hz min$^{-1}$) over that period were 6.1, 11.1 and 13.2, for the S1, S1-δ and S1-o subunits, respectively. This indicates that while the final mass density on 11MUN SAM was slightly higher for the S1-δ subunit than for the S1-o subunit, the S1-o subunit adsorption kinetics demonstrated that its interaction with this specific surface was almost 20% faster than that of the S1-δ subunit.

On the TDT-functionalized surface the highest mass density was shown by the S2 subunit, a behavior distinct from all other surfaces. This high mass density correlates with the highest degree in its frequency decrease, as shown in **Figure 3B**. The adsorbed mass density of this subunit calculated for all other surface modifications was relatively low (**Figure 3B**), especially for FOT-functionalized and 11A1U-functionalized surfaces.

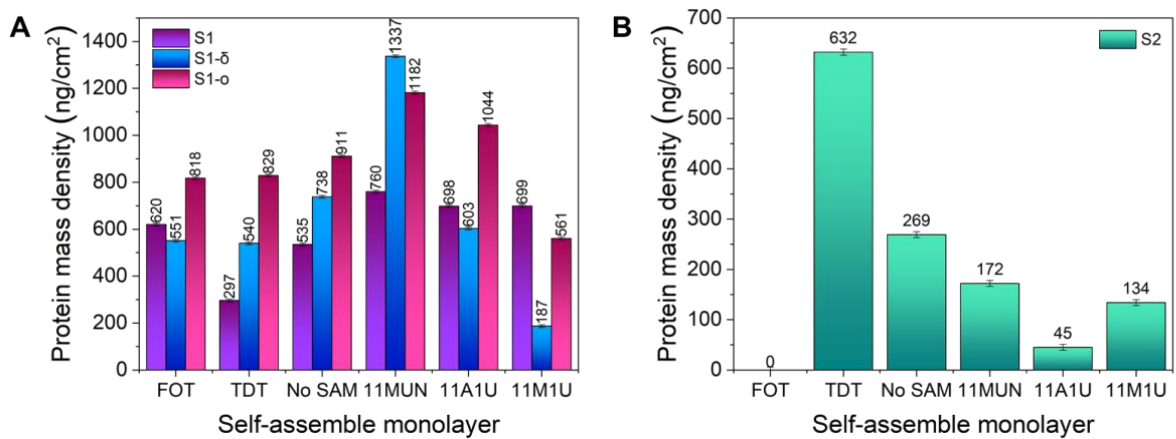

**Figure 3**. Calculated mass density of S1, S1-δ, S1-o (A) and S2 (B) protein layers adsorbed onto various SAM-modified surfaces using Sauerbrey or Voigt models.

**Figure S1** presents the thicknesses of the adsorbed protein layers, calculated by applying the Voigt model or Equation 2, depends on how the mass density was calculated.

$$t = \frac{\Delta m}{\rho} \quad (2)$$

Where *t* is the thickness and ρ is the density of the layer, which can be assumed to be 1 g/cm$^3$.

In light of the stronger binding of the S1 subunits to the 11MUN-functionalized surface, we further studied this adsorption under different pH conditions. **Figure 4** presents changes in frequency and dissipation at different pH values. Using 0.1 M HCl or NaOH 0.1 M solutions, we adjusted protein-free PBS buffer and protein solution to pH values of 8.5, 7.4, 6.5 and 3. The greatest decrease in the frequency was found in the case of pH 7.4 and 6.5 (**Figure 4A**). Interestingly, dissipation was relatively high at pH 6.5, resulting in an adsorbed protein mass as low as that obtained at pH 3 (**Figure 4D**), where the decrease in frequency was lowest.

Overall, the highest amount of S1 adsorption was observed at pH 7.4, while both the more basic and the more acidic pH conditions inhibited adsorption. In addition, plotting of the change in $\Delta D$ as a function of the change in frequency ($|\Delta f|$) enabled us to characterize the viscoelastic properties of the protein layers (**Figure 4C**).

All the adsorbed protein layers remained relatively rigid for the $|\Delta f|<20$, since the slope of the curves (i.e., the change in dissipation) within this range is close to zero. As the protein continued to adsorb, the plots corresponding to pH values 7.4 and 6.5 demonstrated a further increase in slope, indicating that the films were becoming less rigid and more flexible and hydrated, and revealing that at pH 6.5 the protein layer was more viscoelastic than that at pH 7.4. The layers formed at pH 8.5 or 3 revealed only a slight increase in slope over time, suggesting that the adsorbed layers can be considered as rigid and dehydrated layers, and hence that the Sauerbrey equation can be safely applied.[68,69]

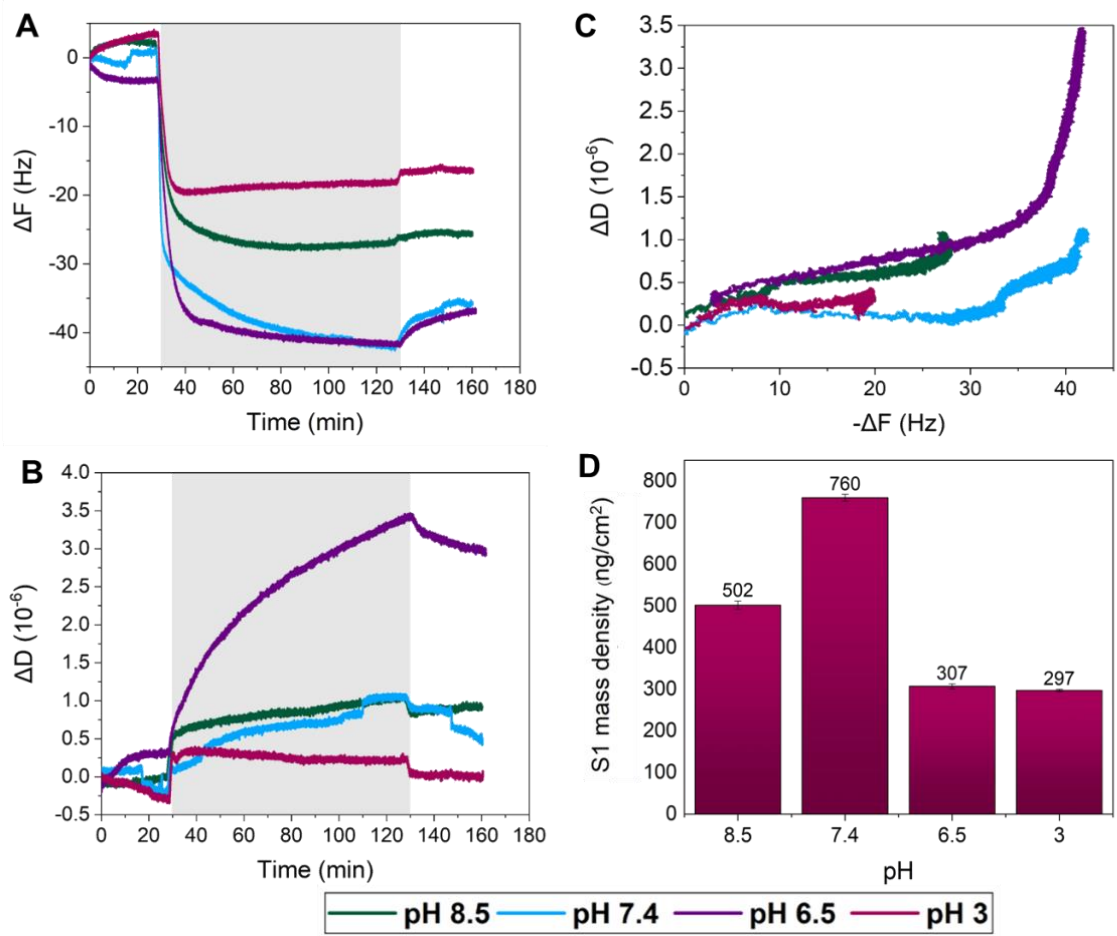

**Figure 4**. (A) Frequency and (B) dissipation curves (3$^{rd}$ overtone) over time at different pH conditions. The grey background represents the adsorption process of the S1 subunit (0.05

mg/mL) and the white background corresponds only to the PBS buffer. (C) $\Delta D \: vs. \: \Delta f$ plot. (D) Calculated mass density of S1 adsorbed layer using Sauerbrey and Voigt models.

3. **Discussion**

Protein adsorption onto surfaces is a complex process that can be affected by several types of interactions (e.g. hydrophobic, electrostatic, ionic) or via hydrogen bonding. Mass and thickness of the adsorbed protein layer are important players in determining the orientation of the proteins on surfaces, and can indicate whether the 3D structure of the protein is preserved or if the protein undergoes denaturation during the process.

As observed from our experiments, the S1 subunits adsorb better onto hydrophilic surfaces such as 11M1U and 11A1U than onto hydrophobic ones. This preferential adsorption can be attributed to the location of the S1 subunit on the periphery of the S protein. With regard to the effect of pH, the S1 subunit adsorbs to a higher extent to the 11MUN SAM surface at a pH of 7.4 than 3, 6.5 and 8.5. At this pH, the 11MUN SAM functional carboxylic acid end group ($pK_a = 5.5-6.0$[70]) undergoes deprotonation and becomes negatively charged. It was previously calculated that the isoelectric point (pI) of the RBD of the S1 subunit is >7.4,[71] and thus most of its surface is positively charged at pH 7.4. This might contribute, by electrostatic interactions, to the strengthening of its adsorption to a negatively charged surface. These results are in line with the fact that the ACE2 receptor which binds RBD is also known to be negatively charged.[72]

The S1-δ subunit showed a significantly higher capacity for adsorption on 11MUN-functionalized surfaces than that of the original SARS CoV-2 S1 subunit, and also more than that of S1-o. This can be explained by considering 9 of the mutations that developed in the S1-δ protein.[73] Four of these mutations (T19R, L452R, T478K, P681R)[73] exhibited a change from a hydrophobic or uncharged amino acid to a positively charged one (arginine or lysine). Those results suggested that the L452R, T478K and P681R mutations indeed have significant impacts on the increase in viral transmissibility,[74] and that this might be the main factor explaining the difference in their adsorption compared to that of the original S1 subunit. In two other mutations, negatively charged amino acids were replaced, one by an uncharged amino acid (D614G) and one was omitted (E156del). In other words, there was an increase in the positive electrostatic potential of the S1-δ subunit.[75] That increase was probably responsible for its significant increase in the ability of this subunit to become adsorbed onto the negatively charged 11MUN-functionalized surface at pH 7.4. This likelihood correlates very well with the reported increase in affinity of the S1-δ subunit to the negatively charged ACE2 receptor, which probably leads to the higher infection rates than those of the nonmutated S1 subunit.[74]

It is a challenging task to understand and predict the influence of the possible mutations potentially present in the S1-o subunit of SARS CoV-2 on their adsorption behaviors on different surfaces. Current mutations in the S1-o subunit probably lead to a 3D structural change that results in a less effective interaction than that of the original S subunit with the TMPRSS2 (which is abundant in the lungs), and therefore causes a less severe disease.[76] Previous studies showed that the electrostatic potential of the S1-o subunit becomes even more strongly positive than that of the S1-δ subunit, especially in the RBD,[77] and this might be one of the reasons for the increase in infection rates by the Omicron variant globally. It might be expected that adsorption of the S1-o subunit onto an 11MUN-functionalized surface would be higher than that of the S1-δ subunit, but this is not necessarily the case, and an observed change in adsorption capacity might be due to other structural changes caused by the mutations. Previous studies have shown that the affinity of the RBD for the ACE2 receptor from S1-o is weaker than the RBD from S1-δ, and stronger than the RBD from the original S1 subunit.[78,79] This correlates well with the adsorption rates we obtained using 11MUN. It is important to note that when we consider the average mass density adsorption, the Omicron variant shows the highest overall adsorption capacity to various types of surfaces. This means that the S1-o subunit can create more pronounced chemical bonds with different surfaces, which is more realistic for real-life surfaces.

All S1 subunits also demonstrate capabilities for adsorption onto hydrophobic SAM surfaces such as TDT and FOT, as well as onto the bare gold surface. This observation suggests that during the adsorption process some conformational changes probably occur in the protein: namely, that specific subunit domains become exposed at the surface, and that this can lead to an increase in their adsorption capacity onto hydrophobic substrates. **Figure 1B** shows the distribution of hydrophilic and hydrophobic amino acids on the surface of the S protein. Even though the surface of most of the S1 subunit contains hydrophilic amino acids (top view), there are also hydrophobic regions that might contribute to this adsorption process onto hydrophobic surfaces. It is interesting to note that in comparison to S1, both S1-δ and S1-o subunits tend to adsorb more successfully to hydrophobic surfaces. This might suggest that mutations in the S1-δ and S1-o subunits significantly affect the 3D structure and the activity of the S protein.

The length of the whole S protein (vertically to the virus membrane) is ~15 nm[80], and assuming that the S1 and S2 subunits are each about half of this length, we would expect that the monolayer thickness would be approximately 7.5 nm if the orientation of the adsorbed protein is vertical to the surface and closely packed. For the S1 subunit, we observed an adsorbed layer thickness of about 7 nm when using sensors modified with 11M1U, TDT and 11MUN SAMs, which might suggest that the packing is indeed close, and that the orientation of the protein is almost vertical (**Figure S1A**).[81,82] We cannot precisely determine which side of the protein adsorbs onto the sensor, meaning that we cannot establish via this method whether the RBD is oriented towards the surface and interacts with it directly, or is oriented rather towards the outer environment.

On the other hand, the thicknesses on FOT- and on 11A1U-functionalized surfaces and on bare gold are smaller (**Figure S1A**). There are two possible reasons for this: (*i*) low protein density on the surface, leading to non-dense and nonhomogenous packing, or (*ii*) a side-on orientation of the protein on the surface, lowering the thickness of the adsorbed layer. While the first explanation would seem to be more reasonable because of the low mass density, our results can be also explained by a combination of these two phenomena.

The S1-δ and S1-o subunits exhibit significantly thicker layers, of about 14 and 12 nm, respectively, on the 11MUN-functionalized surface than on all other surfaces. S1-o shows relatively high thickness (10.44 nm) on 11A1U-functionalizes surface as well. This might indicate the formation of a protein multilayer rather than a ~7.5-nm single protein monolayer. This finding emphasizes the high affinity of the S1-δ or the S1-o subunit for the 11MUN-modified surface.

As can be seen in **Figure 3B**, the S2 subunit adsorbed most strongly to the hydrophobic TDT-functionalized surface. This can probably be attributed to its hydrophobic nature, shown in **Figure 1B** (rear view), and it correlates well with the transmembrane location of the S2 subunit. The thickness of the adsorbed proteinous layer on the TDT-functionalized surface was about 6.5 nm (**Figure S1B**), which might also suggest a close packing arrangement, since it approximates the full length of the S2 subunit. Substantially lower adsorption was measured on 11A1U- and FOT-functionalized surfaces, and as expected, the thickness of the proteinous layer was close to zero (**Figure 3D**). These results are in good agreement with previous reports showing low adhesion of proteins and other macromolecules to perfluorinated surfaces.[83] However, since all S1 subunits, and especially S1-o, demonstrate high affinity to FOT, it indicates that even fluorinated surfaces do not offer a good solution towards inhibiting the attachment of S proteins and reducing the adsorption of the SARS CoV-2 virus.

With regard to the effect of pH, it is known that this factor strongly influences the 3D structure of proteins and their functions. If the pH is altered, with resulting changes to the structure and function of the protein, these conformational changes may be restored when optimal pH conditions are recovered. In some cases, however, the protein undergoes irreversible denaturation. The S protein is stable over a wide range of pH values (3−9), and has been shown to withstand even extreme acidic conditions (pH 2−3) for up to 1 day.[27,30,84]

Characterization of a pH-dependent structural conformation of the S protein showed that the pH greatly affects the orientation of the RBD domain.[85,86] It was reported that if the RBD is in an open state (also termed "up state") it is capable of bonding with the ACE2 receptor, whereas in the closed state (also called "down state") such interaction is limited.[87–89] Interestingly, it was proved that the "open state" is better stabilized at the physiological pH of 7.4 (64% − 68%) than at a pH of 6.5 (44% − 46%) or 8 (39% − 41%).[90] This probably explains our results showing that S1 adsorption is clearly more pronounced at pH 7.4, whereas under either more acidic or more basic conditions there is a decrease in the adsorption level. Since the RBD is positively charged, we can assume that its "open state" allows adsorption in the case where the RBD is oriented toward a negatively charged surface. Reducing or increasing the pH will probably result in a conformational change, as well as a change in the RBD to a "closed state". This might explain the decrease in adsorption onto the 11MUN SAM surface at pH values of 6.5 and 8.5, owing to a less surface-exposed, positively charged RBD.

Another important factor to consider when studying the effect of pH is the pI of the whole S1 subunit within the range of 7.68 to 8.13,[71] and also that the net charge of the S1 subunit is positive at pH values <pI and negative at pH >pI. Therefore, when comparing the charge of the 11MUN SAM surface and that of the S1 subunit under various pH conditions it is important to understand the role of electrostatic interactions in the adsorption to these surfaces. At pH 3 the carboxylic acid end group of the SAM is protonated, meaning that the surface is neutral, and we can indeed obtain the lowest adsorption of the S1 subunit (297 ng cm$^{-2}$, **Figure 4D**). In the range of pH 6.0−7.68, where the S1 subunit is positively charged and the surface is negative, we obtain the highest adsorption specifically at pH 7.4. As a comparison, however, at pH 6.5 the 11MUN carboxylic acid is close to its pK$_a$ value and therefore becomes less negative, and we indeed see a significant decrease in adsorption (307 ng cm$^{-2}$**, Figure 4D**). The adsorption mass density at pH 6.5 is similar to that at pH 3, implying that pH values lower than 6.5 do not affect the interaction between the S1 subunit and the surface.

Given the fact that the SARS CoV-2 virus attaches first to mucosal tissues in an individual's airway, it is interesting to note that the surface of mucosal tissues (as opposed to the skin) is coated by non-keratinized epithelium.[91] The multifunctional nature of mucin allows various types of interactions, as reflected by its structure and unique chemistry. The central core of the mucin molecule is thiol-rich and contains both hydrophobic and charged hydrophilic domains. The glycosylated regions enable intermolecular hydrogen bonding which, under physiological pH conditions, confer a net negative charge due to the presence of ester sulfate ($pK_a = 1.0$) and sialic acid ($pK_a = 2.6$) residues that favor electrostatic interactions with positively charged molecules and particulate matter, including nanostructures such as viruses and nano-pollutants.[92]

Our results provide strong evidence for the fundamental role of electrostatic interactions in the adsorption of SARS-CoV-2 to surfaces. This, together with our knowledge of the structure and properties of mucus (the outermost layer of every mucosal tissue), is corroborated by the rapid invasion of the respiratory system shown by this virus globally.

## 4. Conclusions

In this work, using QCM-D, we compared the adsorption behaviors of 4 different proteins (S1, S1-δ, S1-o and S2 subunits originating from SARS-CoV-2) on chemically modified surfaces comprising SAMs with different hydrophobic/hydrophilic characteristics. Comparison of the adsorption of the S1 subunits from the different strains revealed that they behave differently in the adsorption process. In all S1 subunits the adsorption was highest to the negatively charged 11MUN SAM, but the S1-δ and S1-o subunits showed much higher adsorption than the S1 subunit, probably because their positive electrostatic surface charges were higher than that of the original S1 subunit. These findings correlate well with the enhanced infectivity of the Delta and Omicron variants observed globally.

The S1-o subunit demonstrated the highest capacity for adsorption to the various surfaces, which might corroborate the finding that its infectivity was the highest among all other variants. The finding that the S2 subunit adsorbed best to the hydrophobic SAM TDT can probably be explained by that subunit's structure and function and the hydrophobic amino acids exposed on its surface. The best adsorption of the S1 subunit to 11MUN SAM occurred at pH 7.4, which is the physiological pH, and both more acidic and more basic pH values caused a significant decrease in adsorption. These findings might help us to better understand the process of adsorption onto different surfaces under various pH conditions, and may also be applicable to the process of attachment to the ACE2 receptor during infection. Future studies should address the adsorption capacities of whole SARS CoV-2 viruses to various surfaces and compare the results with those of the current study, which relates to the adsorption capacities of the spike protein.

## 5. Materials and methods

### 5.1. Materials

Hydrogen peroxide solution ($H_2O_2$, 30%) was purchased from Sigma-Aldrich (St. Louis, MO, USA), and ammonium hydroxide, absolute ethanol, concentrated HCl and NaOH solutions from Bio-Lab Ltd. (Jerusalem, Israel). SAM surfaces were formed utilizing 1-tetradecanethiol (TDT), 11-mercaptoundecanoic acid (11MUN), 11-mercapto-1-undecanol (11M1U), 3,3,4,4,5,5,6,6,7,7,8,8,8-tridecafluoro-1-octanethiol (FOT) (all supplied by Sigma-Aldrich), and 11-amino-1-undecanthiol (11A1U) from GERBU Biotechnik GmbH (Heidelberg, Germany). Hellmanex™ III was supplied by Sigma-Aldrich.

Recombinant SARS-CoV-2 S1 and S2 subunits were purchased from R&D Systems (Minneapolis, MN, USA), S1-δ from BPS Bioscience (San Diego, CA, USA) and S1-o from Genscript Biotech (Piscataway, NJ, USA). Sterile phosphate-buffered saline (PBS, pH 7.4) was from Sigma-Aldrich.

### 5.2. Water contact angle

Water contact angle measurements were performed with an Attension Theta Tensiometer (KSV Instruments Ltd., Espoo, Finland). For this purpose, each sensor was measured 3−5 times with 7 μL deionized (DI) water droplets. Results are expressed as mean values and standard deviations.

### 5.3. Preparation of the QCM-D

QCM-D gold-coated sensor crystals with a fundamental frequency of 5 MHz were purchased from Renlux Crystal Ltd. (Shenzhen, China). The sensors were cleaned in 3 steps according to the Q-Sense official protocol:[66] (i) exposure to UV light (10 min), (ii) immersion in piranha solution containing deionized water (DIW), $H_2O_2$ 30%, and ammonium hydroxide in a volume ratio of 5:1:1 (5 min, 75°C), and washing with DIW and drying with $N_2$, and (iii) exposure to UV light (10 min). To change the chemical properties of the sensor we used thiol SAMs, known for their great affinity to gold and their ability to form well-ordered monolayers. To coat the sensors with different SAMs, the relevant SAM was dissolved in absolute ethanol to form a concentration of 5 mM (except for 11A1U, which was prepared to a concentration of 1 mM), and the gold-coated sensor was immersed within the SAM solution for 24 h. The sensor was then immersed in absolute ethanol to remove unbound SAM and dried with $N_2$ gas. Each crystal sensor was used only once. Proteins were dissolved in degassed PBS at a final concentration of 0.05 mg/mL and used in a closed cycle utilizing a peristaltic pump. Since the adsorbed mass after each experiment was equal to or less than 1% of the total amount, we assumed that the protein concentration remained constant throughout the experiment. After each experiment the QCM system was cleaned with 20 mL of Hellmanex™ III (4% solution), and then immersed in DIW (100 mL).

### 5.4. Monitoring Protein Adsorption Using the QCM-D System

QCM-D measurements were performed with the Q-SENSE E1 system (Q-Sense AB, Gothenburg, Sweden) and a peristaltic pump (Izmatec, Wertheim, Germany). The QCM-D chamber was maintained at 20.0 ± 0.1 °C and the flow rate during the experiments was set to 0.1 mL/min. In all experiments the sensor was placed in the QCM for overnight stabilization. Each measurement was first stabilized for 30 min with a protein-free PBS pH 7.4 buffer solution to obtain a stable baseline. Once this was achieved, the desired protein solution was added for 100 min. Measurement data for $f$ and $D$ were acquired at all possible harmonics (n = 1−13) simultaneously. For the Sauerbrey model, the third overtone (n = 3) was presented and used for calculations. The third to eleventh overtones (n = 3, 5, 7, 9 and 11) were used with the viscoelastic Voigt model for soft films, using "QTools" software (Q-Sense AB). Solution viscosity and density were set as fixed parameters, which were assumed to be 1000 kg m$^{-3}$ and 0.001 kg ms$^{-1}$, respectively.

# Supporting Information

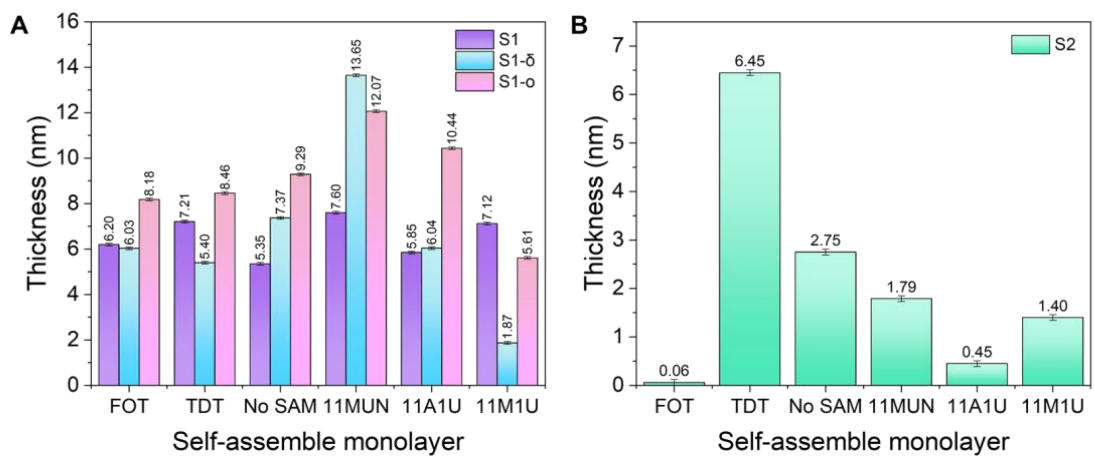

**Figure S1**– Calculated thickness of S1, S1-δ, S1-o (A) and S2 (B) protein layers adsorbed onto various SAM-modified surfaces using Equation 2 and Voigt model.

## Acknowledgements

This research was funded by the Israel Science Foundation (ISF Grant 3970/19). B.P. thanks the Bank Discount Academic Chair for financial support. A.S. thanks the Tamara and Harry Handelsman Academic Chair for financial support. B.P. thanks Prof. A. Futerman from The Weizmann Institute of Science and Dr. E. B. Vitner from the Israel institute for Biological Research for helpful discussions. We are also grateful to E. Stein from the Technion for helpful advice on conducting the experiments.